\documentclass[final,5p,times,twocolumn]{elsarticle}

\usepackage{graphicx}

\usepackage{amsmath}%
\usepackage{amsfonts}%
\usepackage{amssymb}%
\usepackage[usenames]{color}%

\begin{document}

\begin{frontmatter}



\title{Comments on `` Translation-invariant bipolarons and the problem
of high-temperature superconductivity''}


\author[1,2]{S. N. Klimin}
\author[1,3]{J. T. Devreese\corref{cor1}}
\cortext[cor1]{Phone: +32-3-2652485 \quad Fax: +32-3-2653318}
\ead{jozef.devreese@ua.ac.be}
\address[1]{Theorie van Kwantumsystemen en Complexe Systemen (TQC), Universiteit
Antwerpen, Universiteitsplein 1, B-2610 Antwerpen, Belgium}
\address[2]{Department of Theoretical Physics, State University of
Moldova, MD-2009 Chisinau, Moldova}
\address[3]{COBRA, Eindhoven University of Technology, 5600 MB Eindhoven,
The Netherlands}

\begin{abstract}
We comment on the recent results of Refs. \cite{L1,L2} on the bipolaron
problem derived using an approximation of Gross -- Tulub. It is proved that,
contrary to the claim made in Refs. \cite{L1,L2}, the bipolaron ground state
energy calculated there in the strong-coupling approximation has not been
shown to constitute a variational upper bound.
\end{abstract}


\end{frontmatter}





\section{Introduction}

Fr\"{o}hlich-bipolarons play a role in the study, e.g., of electronic and
magnetic properties of polar solids
\cite{R1,R2,Alexandrov,Verbist1991,V2,Bassani,Adamowski,Iadonisi1,Iadonisi,Calvani,Alex1}%
. Bipolarons have, e.~g., been invoked in studies of high-$T_{c}$
superconductivity \cite{Alexandrov,Alex1}. It was shown in
\cite{Verbist1991,V2} (see also Ref. \cite{R1}) that bipolarons can exist only
in a highly restricted stability-region of the $\left(  U,\alpha\right)  $
plane (where $\alpha$ is the electron-phonon coupling constant and $U$ is the
strength of the Coulomb repulsion) and that some of the high-$T_{c}$ oxides
belong to this stability region, whereas conventional polaron-materials
(alkali-halides, Silver-halides etc\ldots) have $\left(  U,\alpha\right)  $
parameters far away from the bipolaron stability region. Bipolarons are only
stable at sufficiently large $\alpha$, so that intermediate and strong
coupling are relevant for the present discussion. Importantly for the study of
high-$T_{c}$ superconductivity, it was found \cite{Verbist1991,V2} that a
bipolaron binds more easily in 2D than in 3D and that the average pair-radius
is a few Angstroms. The groundstate properties of bipolarons were studied
further, e. g., in Refs. \cite{Bassani,Adamowski,Iadonisi1,Iadonisi}.

Recently, the large-bipolaron problem was approached in Refs. \cite{L1,L2}
using the Gross -- Tulub (GT) approximation \cite{Gross,Tulub} in the
strong-coupling limit. Surprisingly low upper bounds to the bipolaron
groundstate energy were arrived at in \cite{L1,L2}. In the limiting case for
the ratio of the dielectric constants $\eta\equiv\varepsilon_{\infty
}/\varepsilon_{0}=0$ and the Fr\"{o}hlich coupling constant $\alpha\gg1$, the
bipolaron groundstate energy proposed as an upper bound in \cite{L2} is
$E_{bip}\left(  \alpha\gg1,\eta=0\right)  \approx-0.414125\alpha^{2}$,
significantly lower than any other result in the literature.

In the present communication we analyse the method and the results of Refs.
\cite{L1,L2} and we demonstrate that the approximation for the bipolaron
strong coupling groundstate energy arrived at in \cite{L2} has not been shown
-- contrary to the claim in \cite{L2} -- to constitute an upper bound.

\section{General treatment in Refs. \cite{L1,L2}}

Consider the two-polaron (bipolaron) system with the Hamiltonian%
\begin{align}
\hat{H}  & =-\frac{\hbar^{2}}{2m}\Delta_{1}-\frac{\hbar^{2}}{2m}\Delta
_{2}+\frac{e^{2}}{\varepsilon_{\infty}\left\vert \mathbf{r}_{1}-\mathbf{r}%
_{2}\right\vert }\nonumber\\
& +\sum_{\mathbf{k}}\hbar\omega_{0}a_{\mathbf{k}}^{\dag}a_{\mathbf{k}%
}\nonumber\\
& +\sum_{\mathbf{k}}V_{\mathbf{k}}\left(  a_{\mathbf{k}}+a_{-\mathbf{k}}%
^{\dag}\right)  \left(  e^{i\mathbf{k\cdot r}_{1}}+e^{i\mathbf{k\cdot r}_{2}%
}\right)  .\label{Hstart}%
\end{align}
with the coupling parameters
\begin{equation}
V_{\mathbf{k}}=\frac{\hbar\omega_{0}}{k}\left(  \frac{2\sqrt{2}\pi\alpha}%
{V}\right)  ^{1/2}\left(  \frac{\hbar}{m\omega_{0}}\right)  ^{1/4}.
\end{equation}
Here, $\alpha$ is the Fr\"{o}hlich polaron coupling constant, $\varepsilon
_{\infty}$ is the high-frequency dielectric constant. The system of units is
chosen with $2m=1$, $\hbar=1$, and the LO-phonon frequency $\omega_{0}=1$.
After the transformation to the center-of-mass and relative coordinates%
\begin{equation}
\mathbf{R}=\frac{\mathbf{r}_{1}+\mathbf{r}_{2}}{2},\;\mathbf{r}=\mathbf{r}%
_{1}-\mathbf{r}_{2},
\end{equation}
a first Lee-Low-Pines (LLP)-type canonical transformation%
\begin{equation}
\hat{S}_{1}=\exp\left(  -i\mathbf{R}\cdot\sum_{\mathbf{k}}\mathbf{k}%
a_{\mathbf{k}}^{\dag}a_{\mathbf{k}}\right)  ,\label{S1}%
\end{equation}
and the averaging of the transformed Hamiltonian with a trial wave function
$\varphi\left(  r\right)  $ for the relative motion (in the oscillatory form
as in Refs. \cite{L1,L2}),
\begin{equation}
\varphi\left(  r\right)  =\frac{1}{\left(  \pi\rho^{2}\right)  ^{3/4}}%
\exp\left(  -\frac{r^{2}}{2\rho^{2}}\right)  \label{osc}%
\end{equation}
we arrive at the reduced Hamiltonian%
\begin{align}
\tilde{H} &  =\frac{1}{2}\left(  \sum_{\mathbf{k}}\mathbf{k}a_{\mathbf{k}%
}^{\dag}a_{\mathbf{k}}\right)  ^{2}+\sum_{\mathbf{k}}a_{\mathbf{k}}^{\dag
}a_{\mathbf{k}}\nonumber\\
&  +\sum_{\mathbf{k}}4\left(  \frac{\pi\alpha}{k^{2}V}\right)  ^{1/2}%
e^{-\frac{1}{16}\rho^{2}k^{2}}\left(  a_{\mathbf{k}}+a_{-\mathbf{k}}^{\dag
}\right)  \nonumber\\
&  +\frac{3}{\rho^{2}}+\frac{4\alpha}{\sqrt{\pi}\left(  1-\eta\right)  \rho
}.\label{Hs2}%
\end{align}

The second canonical transformation%
\begin{equation}
S_{2}=e^{-\sum_{\mathbf{k}}f_{\mathbf{k}}\left(  a_{\mathbf{k}}-a_{\mathbf{k}%
}^{\dag}\right)  }\label{S2}%
\end{equation}
with the trial phonon shifts $f_{\mathbf{k}}$ chosen as in Ref. \cite{L2} with
the variational parameter $\mu$,%
\begin{equation}
f_{\mathbf{k}}=-4\left(  \frac{\pi\alpha}{k^{2}V}\right)  ^{1/2}\exp\left(
-\frac{1}{2\mu}k^{2}\right)  ,\label{fk}%
\end{equation}
and the generalized Bogoliubov transformation used in Ref. \cite{Tulub} result
in the following variational functional for the bipolaron groundstate energy%
\begin{align}
E_{bip} &  =E_{R}+\frac{3}{\rho^{2}}+\frac{4}{\sqrt{\pi}}\frac{\alpha}{1-\eta
}\frac{1}{\rho}\nonumber\\
&  +\frac{8\sqrt{2}\alpha}{\sqrt{\pi}}\frac{1}{\sqrt{\rho^{2}+\frac{8}{\mu}}%
}-\frac{16\sqrt{2}\alpha}{\sqrt{\pi}}\frac{1}{\sqrt{\rho^{2}+\frac{4}{\mu}}%
}\label{evar}%
\end{align}
with the parameters%
\begin{equation}
\tilde{\alpha}=4\sqrt{2}\alpha,\quad a=\frac{2\sqrt{2}}{\sqrt{\rho^{2}%
+\frac{8}{\mu}}}.
\end{equation}
and the recoil energy%
\begin{equation}
E_{R}=\frac{3}{16}a^{2}\left[  1+Q\left(  \tilde{\alpha},a\right)  \right]
,\label{ER}%
\end{equation}
which results from the recoil term \cite{Roseler} $\frac{1}{2}\left(
\sum_{\mathbf{k}}\mathbf{k}a_{\mathbf{k}}^{\dag}a_{\mathbf{k}}\right)  ^{2}$
in the Hamiltonian (\ref{Hs2}). The function $Q\left(  \alpha,a\right)  $ is
given by the integral expression \cite{Tulub}%
\begin{equation}
Q\left(  \alpha,a\right)  =\frac{2}{\sqrt{\pi}}\int_{0}^{\infty}%
\frac{e^{-y^{2}}\left[  1-\Omega\left(  y\right)  \right]  dy}{\left[
\frac{1}{\lambda}+v\left(  y\right)  \right]  ^{2}+\frac{\pi}{4}%
y^{2}e^{-2y^{2}}},\label{Q}%
\end{equation}
with the functions%
\begin{align}
v\left(  y\right)   &  =1-ye^{-y^{2}}\int_{0}^{y}e^{t^{2}}dt-\xi\int_{\xi
}^{\infty}e^{\xi^{2}-t^{2}}dt,\label{v}\\
\Omega\left(  y\right)   &  =2y^{3}\left[  \left(  1+2y^{2}\right)  \int
_{y}^{\infty}e^{y^{2}-t^{2}}dt-y\right]  ,\label{W}%
\end{align}
and the parameters%
\begin{equation}
\lambda=\frac{4\alpha}{3\sqrt{2\pi}}a,\quad\xi=\sqrt{y^{2}+\frac{4}{a^{2}}%
}.\label{pars}%
\end{equation}

\section{Using the approximation of Ref.~\cite{Tulub} in Ref. \cite{L2}}

In Ref. \cite{Tulub}, the function $Q\left(  \alpha,a\right)  $ [denoted in
\cite{Tulub} as $q\left(  1/\lambda\right)  $] has been replaced by the value
$Q_{\infty}\approx5.75$. If the recoil energy (\ref{ER}) is calculated using
$Q\left(  \alpha,a\right)  =Q_{\infty}$, one arrives at the result of Ref.
\cite{L2} for the bipolaron energy (Eq. (15) of Ref. \cite{L2}). However, this
approximation does not guarantee a variational upper bound for the polaron and
bipolaron groundstate energies for the following reason.%

\begin{figure}
[h]
\begin{center}
\includegraphics[
height=2.2231in,
width=2.7056in
]%
{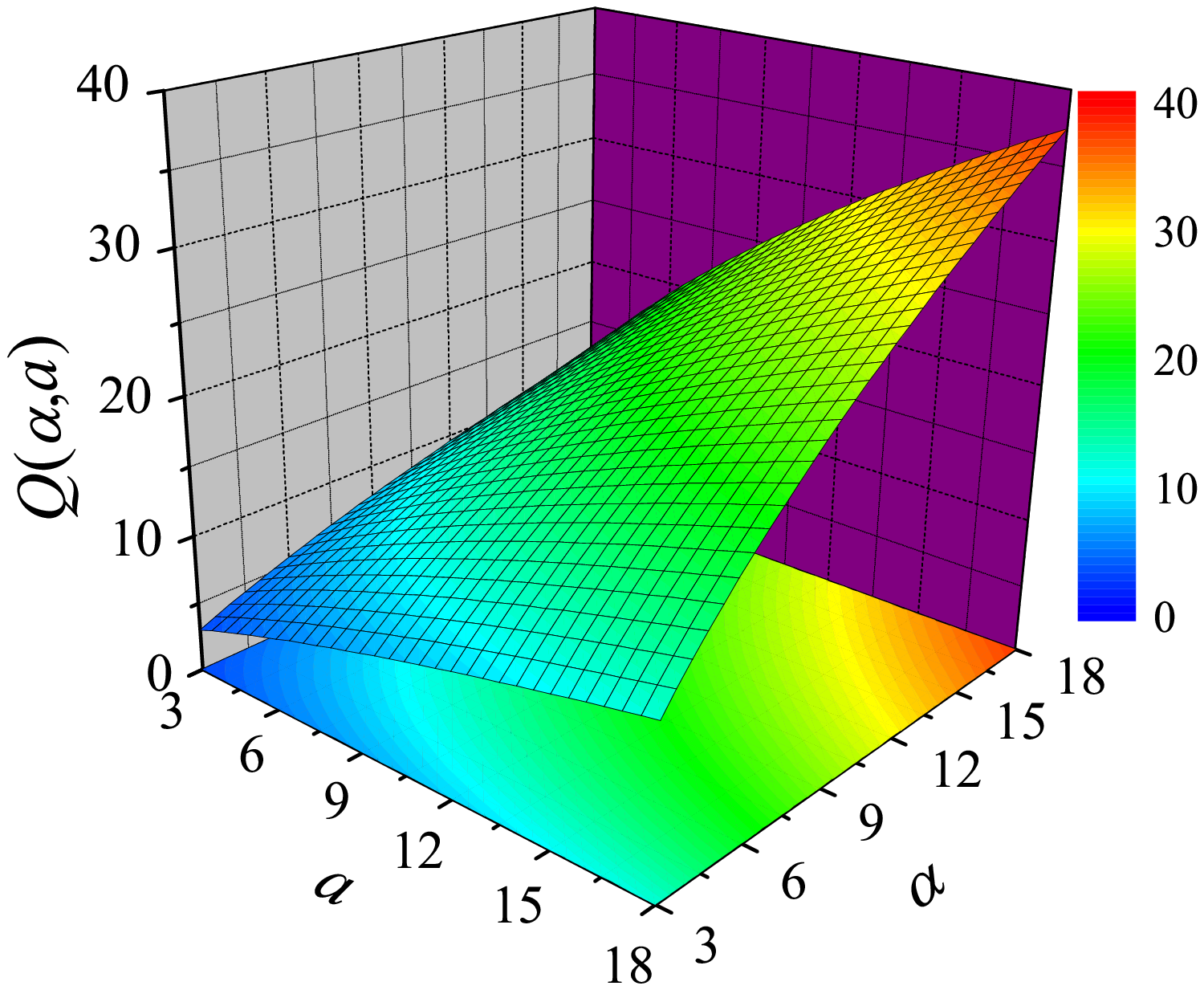}%
\caption{$Q\left(  \alpha,a\right)  $ given by Eq. (\ref{Q})}%
\end{center}
\end{figure}

For finite parameters $\alpha$ and $a$, $Q\left(  \alpha,a\right)  $ is an
increasing function of both $\alpha$ and $a,$ as shown in Fig. 1. Also in the
strong-coupling limit, when both $\alpha$ and $a$ tend to infinity, the
asymptotic expression for this function,%
\begin{equation}
\left.  Q\left(  \alpha,a\right)  \right\vert _{\alpha,a\gg1}\approx
2\sqrt{\frac{2\sqrt{2}}{\sqrt{\pi}}a\alpha},\label{as}%
\end{equation}
increases monotonically. The value $Q_{\infty}\approx5.75$ has been obtained
in Ref. \cite{Tulub} assuming a finite cutoff for the phonon wave vectors.
Indeed, the phonon wave vectors are restricted by a short-wavelength cutoff at
the boundary of the Brillouin zone. Consequently, in the strong-coupling limit
$\alpha\rightarrow\infty$, the position of the steep maximum mentioned in Ref.
\cite{Tulub} \footnote{As written in Ref. \cite{Tulub} on p. 4,
\textquotedblleft It is of interest to note that as $\lambda\rightarrow\infty$
the integrand in (2.12) has a \textit{steep maximum} at $y^{4}=3\lambda/4$;
however, if we take into account that the domain of integration over $y$ is in
fact limited and if we use the values $g^{2}\approx10$ considered in the
following, this singularity does not arise.\textquotedblright\ } can lie
beyond the integration range. However, as far as the calculation of the
(bi)polaron energy is performed within the continuum approach, it is only
consistent either to avoid a cutoff all together (in the continuum formalism),
or to introduce a cutoff from the very beginning of the calculation. A
consistent treatment of the polaron problem in the Gross -- Tulub
approximation using a cutoff was performed in Ref. \cite{Roseler}. It was
found that a cutoff leads to the appearance of additional positive terms in
the polaron recoil energy. These terms were not found in Ref. \cite{Tulub}.
Being positive, they definitely lead to an increase of the groundstate energy.
Therefore missing these terms can lead to the violation of the variational
principle. Hence the bipolaron groundstate energy arrived at in Refs.
\cite{L1,L2} is incorrectly claimed to constitute an upper bound for the
groundstate energy of the bipolaron.

\section{Bipolaron variational energy in the continuum approach}

Let us consistently consider the polaron and bipolaron groundstate energies
within the continuum approximation using the complete recoil energy (i. e,
without a cutoff). In the strong-coupling limit, using (\ref{evar}) with the
asymptotic expression (\ref{as}), the polaron and bipolaron groundstate
energies are then found to be%
\begin{align}
E_{pol}^{\left(  sc\right)  } &  \approx-0.31683\alpha^{4/3},\label{Ep}\\
E_{bip}\left(  \alpha\gg1,\eta=0\right)   &  \approx-0.868509\alpha
^{4/3}.\label{Eb}%
\end{align}
Therefore, when the (bi)polaron groundstate energy is consistently calculated
within the continuum GT approach, an incorrect dependence $E\propto
\alpha^{4/3}$ results in the strong-coupling limit. This problem, however, was
not realised by the author of Refs. \cite{L1,L2}.%

\begin{figure}
[h]
\begin{center}
\includegraphics[
height=2.4132in,
width=3.1871in
]%
{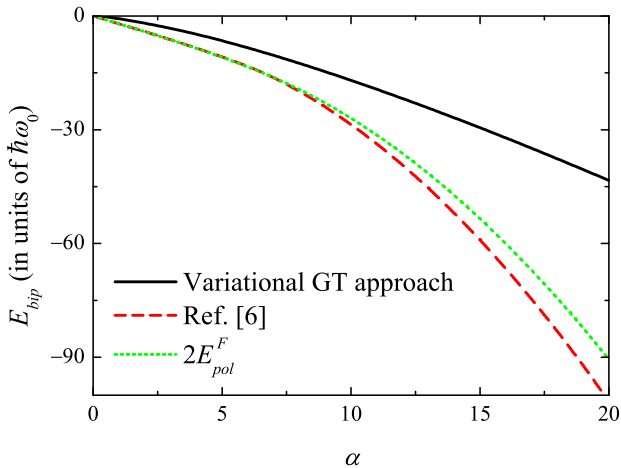}%
\caption{Bipolaron variational groundstate energy calculated using the
complete recoil energy within the Gross -- Tulub scheme in the continuum
approach. For comparison, the bipolaron groundstate energy calculated in Ref.
\cite{Verbist1991} and twice the groundstate energy of a single polaron from
Ref. \cite{Feynman1955} are plotted in the same graph.}%
\end{center}
\end{figure}

In order to check whether the continuum variational GT approach is adequate
for intermediate $\alpha$, we have calculated the bipolaron groundstate energy
using the expression (\ref{evar}) with $Q\left(  \alpha,a\right)  $ given by
(\ref{Q}) instead of the value $Q_{\infty}\approx5.75$ used in Refs.
\cite{L1,L2}. In Fig. 2, this bipolaron groundstate energy is compared with
the variational result of Ref. \cite{Verbist1991} and with twice the energy of
a single polaron calculated using the path-integral variational method of
Feynman \cite{Feynman1955}. It is seen that for intermediate $\alpha$, the
bipolaron groundstate energy calculated using the variational GT scheme in the
continuum approach lies above twice the energy of a single polaron.

\section{Conclusion}

In the present communication we have proved that the strong-coupling
expression for the bipolaron groundstate energy calculated in Refs.
\cite{L1,L2} contrary to what is claimed in those works is not justified as
upper bound for the bipolaron groundstate energy. A contribution to the recoil
energy due to a momentum cutoff \cite{Roseler} within the GT scheme is missed
in \cite{L1,L2}. Thus the results of Refs. \cite{L1,L2} have been obtained
using the \textit{incomplete recoil energy}.

We have also worked out the GT scheme using the complete recoil energy within
the continuum approach as is necessary for a consistent theory. This leads to
an incorrect dependence $E_{bip}\propto\alpha^{4/3}$ (instead of
$E_{bip}\propto\alpha^{2}$) for the polaron groundstate energy in the
strong-coupling limit. A consistent calculation within the continuum approach,
however, was not performed in Refs. \cite{L1,L2}.

Note also that for intermediate $\alpha$, the continuum GT method leads to a
bipolaron groundstate energy higher than twice the groundstate energy for a
single polaron. Therefore this method fails to describe a bound bipolaron
state for any $\alpha$.




\begin{thebibliography}{00}
\bibitem {R1}In the literature both \textquotedblleft large\textquotedblright%
\ (or \textquotedblleft Fr\"{o}hlich\textquotedblright-) polarons and
\textquotedblleft small\textquotedblright\ (or \textquotedblleft
Holstein\textquotedblright-) polarons are studied. The present communication
addresses Fr\"{o}hlich-polarons.
\end{thebibliography}



\end{document}